\documentclass{article}
\makeatletter
\@addtoreset{equation}{section}

\makeatother

\usepackage{graphics}

\begin{document}

\begin{flushright}
Feb 2007

KUNS-2062
\end{flushright}

\begin{center}

\vspace{5cm}

{\LARGE 
\begin{center}
Non-trivial Tachyon Profiles in Low Energy Effective Theory
\end{center}
}

\vspace{2cm}

Takao Suyama \footnote{e-mail address : suyama@gauge.scphys.kyoto-u.ac.jp}

\vspace{1cm}

{\it Department of Physics, Kyoto University,}

{\it Kitashirakawa, Kyoto 606-8502, Japan }

\vspace{4cm}

{\bf Abstract} 

\end{center}

We study classical solutions of a low energy effective theory of a string theory with tachyons. 
With a certain ansatz, we obtain all possible solutions which are static, weakly coupled and weakly 
curved. 
We find, in addition to the interpolating solutions studied in our previous paper, black hole 
solutions and solutions including the geometry of a capped cylinder. 
Some possible implications of the solutions to closed string tachyon condensation are discussed.

\newpage

\section{Introduction}

\vspace{5mm}

Understanding closed string tachyon condensation is still a challenging problem. 
There are studies on this phenomenon using string field theory \cite{SFT1}\cite{SFT1.5}\cite{SFT2}. 
The study in this direction is promising, because it is necessary to treat a finite value of the 
expectation value of a tachyon for investigations of the properties of the process, and therefore, 
the string perturbation theory is not a powerful tool for this problem. 
However, it seems to be difficult, in practice, to study by this approach with enough accuracy, 
compared with the similar analysis of open string tachyon condensation \cite{Sen}\cite{Schnabl}. 
In fact, a closed string field theory employed in \cite{SFT1}\cite{SFT1.5}\cite{SFT2} 
has a very complicated form. 
Therefore, it is desirable to employ a more simple theory as an approximation of the closed 
string field theory. 
One such candidate is the low energy effective theory of string theory which contains massless 
fields in addition to tachyons, and all the other massive states of the string theory are assumed to 
be integrated out. 
Recently, there are some studies \cite{YZ}\cite{BR}\cite{Suyama} in this direction. 

In this paper, we would like to study classical solutions of the low energy effective theory more 
systematically than our previous investigation \cite{Suyama}. 
We consider static solutions which depend on a single spatial coordinate. 
With a certain ansatz, we determine all classical solutions which are weakly coupled and weakly 
curved so as to be acceptable as low energy descriptions of the system. 
Interestingly, we find classical solutions in which a region, where the tachyon vev grows, is either 
replaced with a black hole or simply eliminated, the latter of which is 
similarly to the phenomenon studied in \cite{Silverstein}. 

This paper is organized as follow. 
In section \ref{EOM}, 
we show the form of the low energy effective theory we study in this paper, and then 
show our ansatz for the fields and simplify the equations of motion. 
In section \ref{general}, we investigate some general properties of the classical solutions. 
We also explain how to solve the equations of motion. 
Then we show some solutions with non-trivial tachyon profiles in section \ref{non-trivial}, 
and study their 
properties. 
In section \ref{mass}, we determine the mass difference among the solutions to discuss their stability. 
Section \ref{discuss} is devoted to discussion. 
The details on how to determine the form of the effective action are explained in 
Appendix \ref{action}. 
Appendix \ref{trivial} contains all possible solutions with a constant $T$.

\vspace{1cm}

\section{Equations of motion} \label{EOM}

\vspace{5mm}

We would like to study possible solutions of the equation of motion of a string theory which contains 
a tachyon in its mass spectrum. 
Specifically, the string theory is assumed to be compactified on a compact manifold $M$, and the 
tachyon corresponds to a relevant operator of a CFT describing $M$. 
To study the solutions, 
we employ a low energy effective action of the string theory, and study its classical 
solutions, regarding them as approximate solutions to the full equation of motion. 
The effective action describes the dynamics of the metric $g_{\mu\nu}$, the B-field $B_{\mu\nu}$, 
the dilaton $\Phi$ and the tachyon $T$. 
All the other massive states are assumed to be integrated out. 
In addition, we assume that all the above fields vary slowly in the spacetime. 
This assumption allows us to ignore terms which contain more than two derivatives 
in the effective action. 
The most general form of the effective action is as follows: 
\begin{equation}
S = \frac1{2\kappa^2}\int d^Dx\sqrt{-g}\ e^{-2\Phi}\Bigl[ R+4(\nabla\Phi)^2-\frac1{12}e^{f(T)}H^2
 -(\nabla T)^2-2V(T) \Bigr]. 
    \label{EA}
\end{equation}
Here we assumed a relation between the equations of motion of this action and 
beta-functionals for the conformal invariance of the underlying worldsheet theory. 
See Appendix \ref{action} for the details. 

Some classical solutions of this action are investigated in \cite{YZ}\cite{BR}\cite{Suyama}. 
In the case $B_{\mu\nu}=0$, the explicit form of $f(T)$ is not necessary for the study of the 
solutions, and therefore, this simple action (\ref{EA}) should properly 
describe the dynamics of this system, as long as the fields are slowly varying. 

\vspace{5mm}

In the following, we study classical solutions of the action (\ref{EA}) including the B-field. 
The equations of motion are 
\begin{eqnarray}
R_{\mu\nu}+2\nabla_\mu\nabla_\nu\Phi-\frac14e^{f(T)}
 H_{\lambda\rho\mu}H^{\lambda\rho}{}_\nu-\nabla_\mu T
 \nabla_\nu T &=& 0, \\
\nabla^2\Phi-2(\nabla\Phi)^2-V(T)+\frac1{12}e^{f(T)}H^2 &=& 0, \\
\partial_\rho(\sqrt{-g}\ e^{-2\Phi+f(T)}H^{\rho\mu\nu}) &=& 0, 
   \label{eomB} \\
\nabla^2T-2\nabla\Phi\cdot\nabla T-V'(T)-\frac1{24}f'(T)e^{f(T)}H^2 &=& 0, 
\end{eqnarray}
where $V'(T)=\partial_TV(T)$ etc. 
We consider the ansatz for the metric 
\begin{equation}
ds^2 = -e^{-2A(x)}dt^2+dx^2+e^{-2B(x)}dy^2+h_{kl}(z)dz^kdz^l, 
\end{equation}
and for the H-flux 
\begin{equation}
H^{txy} = H_0e^{A+B+2\Phi-f(T)}, 
\end{equation}
where $H_0$ is an arbitrary constant and the other components are set to be zero. 
We also assume that $\Phi$ and $T$ are functions of $x$. 
This ansatz is also employed in \cite{Suyama}. 
Then the equation (\ref{eomB}) is solved, and the other equations of motion reduce to 
\begin{eqnarray}
A''+B''-(A')^2-(B')^2+2\Phi''+\frac12H_0^2e^{4\Phi-f(T)}-(T')^2 &=& 0, \\
A''-(A'+B'+2\Phi')A'+\frac12H_0^2e^{4\Phi-f(T)} &=& 0, \\
B''-(A'+B'+2\Phi')B'+\frac12H_0^2e^{4\Phi-f(T)} &=& 0, \\
\Phi''-(A'+B'+2\Phi')\Phi'-V(T)-\frac12H_0^2e^{4\Phi-f(T)} &=& 0, \\
T''-(A'+B'+2\Phi')T'-V'(T)+\frac1{4}f'(T)H_0^2e^{4\Phi-f(T)} &=& 0. 
\end{eqnarray}
The metric components $h_{kl}$ must be Ricci-flat. 
We choose $h_{kl}=\delta_{kl}$ for simplicity. 

As is pointed out in 
\cite{Suyama}, the use of a combination $K=A'+B'+2\Phi'$ simplifies the analysis of 
solutions. 
Due to the equations of motion, the field $K$ satisfies 
\begin{equation}
K'-K^2-2V(T) = 0. 
\end{equation}

For simplicity, we set $f(T)=0$ from now on. 
Then, the fields $T$ and $K$ are decoupled from the other fields, and they are determined by 
\begin{eqnarray}
T''-KT'-V'(T) &=& 0, 
   \label{eomT} \\
K'-K^2-2V(T) &=& 0. 
   \label{eomK}
\end{eqnarray}
The other fields $\Phi$ and $A-B$ are then determined by the equations 
\begin{eqnarray}
\Phi''-K\Phi'-V(T)-\frac12H_0^2e^{4\Phi} &=& 0, 
   \label{eomPhi} \\
(A-B)''-K(A-B)' &=& 0, 
   \label{A-B}
\end{eqnarray}
using the solutions $T(x)$ and $K(x)$ of (\ref{eomT}) and (\ref{eomK}) as known functions. 
The solution must satisfy the constraint 
\begin{equation}
K^2-(A')^2-(B')^2-(T')^2+2V(T)+\frac12H_0^2e^{4\Phi} = 0. 
   \label{constraint}
\end{equation}

\vspace{1cm}

\section{General properties of solutions} \label{general}

\vspace{5mm}

\subsection{Validity of solutions} \label{validity}

\vspace{5mm}

A classical solution of (\ref{eomT})-(\ref{A-B}) 
may allow a strongly coupled region, or a strongly curved background, in general. 
Solutions of such kinds are not reliable for the analysis of this system, because our action 
(\ref{EA}) is valid only for slowly-varying solutions. 
To exclude the singular solutions from our consideration, we determine 
conditions for a classical solution to be weakly coupled and weakly curved everywhere. 

Consider the curvature invariant  
\begin{eqnarray}
& & g^{\mu\rho}g^{\nu\lambda}R_{\mu\nu}R_{\rho\lambda} \nonumber \\
&=& \Bigl(2A'\Phi'-\frac12e^{4\Phi}\Bigr)^2+\Bigl(2B'\Phi'-\frac12e^{4\Phi}\Bigr)^2
 +(2A'B'+2A'\Phi'+2B'\Phi'-e^{4\Phi})^2, 
    \label{R^2}
\end{eqnarray}
where we used the equations of motion. 
For a solution to be weakly curved, this quantity must be finite. 
We also require that $e^{\Phi}$ is finite, but this requirement does not exclude the behavior 
$\Phi'\to-\infty$ at some region. 
Therefore, it should be checked whether (\ref{R^2}) is finite under the behavior $\Phi'\to-\infty$. 
Suppose that $\Phi'$ diverges at $x=x_\infty$. 
Since each terms in (\ref{R^2}) must be individually finite, 
the fields $A'$ and $B'$ must vanish at $x=x_\infty$. 
Then the equation (\ref{eomPhi}) implies 
\begin{equation}
\Phi'' \sim 2(\Phi')^2, 
\end{equation}
where $V(T)$ is assumed to be bounded around $x= x_\infty$. 
However, this equation contradicts the assumed behavior $\Phi'\to-\infty$. 
Therefore, $\Phi'$ must be finite everywhere for a reliable solution of the effective action. 

If $\Phi'$ vanish at $x=x_\infty$, then either $A'$ or $B'$ is allowed to diverge at $x=x_\infty$, 
while keeping the solution weakly curved. 
Note that, due to the invariance of the equations of motion under the exchange of $A$ and $B$, there 
is always a pair of solutions, consisting of one solution with divergent $A'$, and another one with 
divergent $B'$. 

In summary, a weakly coupled and weakly curved solution is obtained only when $\Phi'$ is finite 
everywhere, and either $A'$ or $B'$ is allowed to diverge where $\Phi'$ vanishes. 

\vspace{5mm}

\subsection{Behavior of $T$ and $K$}

\vspace{5mm}

The equation (\ref{eomT}) has a form familiar in the classical mechanics. 
The field $T$ can be interpreted as a coordinate of a particle moving in the potential $-V(T)$. 
Suppose that $V(T)$ has a local maximum at $T=0$. 
Then there exists a solution in which $T$ evolves toward $T=0$, provided that $K$ is negative. 
What we are interested in are such solutions, that is, the solutions with the behavior 
\begin{equation}
\lim_{x\to+\infty}T(x)=0. 
   \label{limitT}
\end{equation}
We call them non-trivial solutions. 
It would be reasonable to expect that the existence of the non-trivial 
solution has the following implication 
to closed string tachyon condensation. 
A solution, to which the non-trivial solution approaches in the $x\to+\infty$ limit, is regarded as 
the initial background of tachyon condensation. 
The tachyon vev becomes larger as $x$ decreases, and the behavior of the non-trivial 
solution around a finite $x$ is regarded as a background which is a deformation of the initial 
background due to the back-reaction of the non-zero tachyon vev. 
If the non-trivial solution approaches another solution in the $x\to-\infty$ limit, then the limiting 
solution should describe an endpoint of tachyon condensation. 
We will show that there are another cases in which an event horizon is formed at a finite $x$, and 
a further evolution of the tachyon is hidden behind the horizon in view of the observer at 
$x=+\infty$. 

The equation (\ref{eomT}) implies 
\begin{equation}
\frac d{dx}\Bigl[ \frac12(T')^2-V(T) \Bigr] = K(T')^2. 
\end{equation}
Therefore, if 
\begin{equation}
K\le 0 \hspace{5mm} \mbox{for }x\ge x_0
   \label{K<0}
\end{equation}
is satisfied for some $x_0$, then the behavior (\ref{limitT}) is realized. 
This is simply because $T$ feels a friction when $K$ is negative. 
The equation (\ref{eomK}) implies that the behavior 
(\ref{K<0}) is realized only if $V(T)\le0$ around $T=0$. 
In this case, we obtain 
\begin{equation}
\lim_{x\to+\infty}K(x) = -\sqrt{-2V(0)}. 
   \label{limitK}
\end{equation}

In summary, we consider a potential $V(T)$ with the form  
\begin{equation}
V(T) = V_0+\frac12m^2T^2+O(T^3)
\end{equation}
around $T=0$, where $V_0\le0$ and $m^2<0$, and we investigate solutions which behave as (\ref{limitT}) 
and (\ref{limitK}). 
Note that the initial background is either a linear dilaton or an $AdS_3$, 
as is shown in Appendix \ref{trivial},  

\vspace{3mm}

The analysis of the linear perturbation around $T=0$ and $K=-\sqrt{-2V_0}$ can be done easily. 
The resulting behavior of $T$ is 
\begin{equation}
T(x) = C_1e^{\omega_+x}+C_2e^{\omega_-x}, 
\end{equation}
where 
\begin{equation}
\omega_\pm = \frac12(-\sqrt{-2V_0}\pm\sqrt{-2V_0+4m^2}). 
\end{equation}
In the case $-2V_0+4m^2<0$, $T$ exhibits a damped oscillation. 
For an $AdS_3$ background, this condition for the oscillation coincides with the condition for the 
mass squared of $T$ to be 
below the Breitenlohner-Freedman bound \cite{BF}\cite{BF2}, that is, $T$ is indeed tachyonic. 

Due to the oscillation of $T$, all the other fields also oscillate in the $x\to+\infty$ limit. 
This fact will be important when we determine the mass of the classical solutions 
in section \ref{mass}. 

\vspace{5mm}

\subsection{Solutions for $A$, $B$ and $\Phi$}

\vspace{5mm}

Interestingly enough, the equations (\ref{eomPhi}) and (\ref{A-B}) can be solved analytically, 
provided that a solution of the equations (\ref{eomT}) and (\ref{eomK}) is given. 

The equation (\ref{A-B}) can be easily integrated once, and we obtain  
\begin{equation}
A'-B' = Ce^{{\cal K}}, 
   \label{intA-B}
\end{equation}
where 
\begin{equation}
{\cal K} = A+B+2\Phi. 
   \label{cal K}
\end{equation}

The equation (\ref{eomPhi}) can be rewritten as follows, 
\begin{equation}
\Bigl(e^{-{\cal K}}\frac d{dx}\Bigr)^2(A+B)+H_0^2 e^{-2(A+B)} = 0, 
\end{equation}
where we used the equation (\ref{eomK}) and the relation (\ref{cal K}). 
The general solution of this equation is 
\begin{equation}
A(x)+B(x) = \log\left|\sinh\left(\sqrt{2E}(\rho(x)-\rho_0)\right)\right|
 +\frac12\log\left(\frac{H_0^2}{2E}\right), 
   \label{A+B}
\end{equation}
where $E$ and $\rho_0$ are integration constants, and the function $\rho(x)$ is defined as a solution 
of the equation 
\begin{equation}
\frac{d\rho}{dx} = e^{\cal K}. 
\end{equation}
Now, we obtain the general solution 
\begin{eqnarray}
A'(x) &=& \frac12e^{{\cal K}(x)}\left[\sqrt{2E}\coth\left(\sqrt{2E}(\rho(x)-\rho_0)\right)+C\right], 
   \label{solnA'} \\
B'(x) &=& \frac12e^{{\cal K}(x)}\left[\sqrt{2E}\coth\left(\sqrt{2E}(\rho(x)-\rho_0)\right)-C\right], 
   \label{solnB'} \\
\Phi'(x) 
 &=& \frac12\left[K(x)-\sqrt{2E}\ e^{{\cal K}(x)}\coth\left(\sqrt{2E}(\rho(x)-\rho_0)\right)\right]. 
   \label{solnPhi'}
\end{eqnarray}

Note that the general solution in the case $H_0=0$ is 
\begin{eqnarray}
A'(x) &=& C_1e^{{\cal K}(x)}, \\
B'(x) &=& C_2e^{{\cal K}(x)}, \\
\Phi'(x) &=& \frac12K(x)-\frac{C_1+C_2}2e^{{\cal K}(x)}, 
\end{eqnarray}
where $C_1$ and $C_2$ are integration constants. 

It is possible to integrate the above expressions, and, in the case $H_0\ne0$,  we obtain 
\begin{eqnarray}
ds^2 &=& \frac1{\sinh\left(\sqrt{2E}(\rho(x)-\rho_0)\right)}
 \left(-e^{-C\rho(x)}dt^2+e^{C\rho(x)}dy^2\right)+dx^2, \\
\Phi &=& \Phi_0+\frac12{\cal K}-\frac12\log\sinh\left(\sqrt{2E}(\rho(x)-\rho_0)\right), 
\end{eqnarray}
for the case $E>0$, and 
\begin{eqnarray}
ds^2 &=& \frac1{\rho(x)-\rho_0}\left(-e^{-C\rho(x)}dt^2+e^{C\rho(x)}dy^2\right)+dx^2, \\
\Phi &=& \Phi_0+\frac12{\cal K}-\frac12\log(\rho(x)-\rho_0), 
\end{eqnarray}
for the case $E=0$. 
In the case $H_0=0$, we obtain 
\begin{eqnarray}
ds^2 &=& -e^{-2C_1\rho(x)}dt^2+dx^2+e^{-2C_2\rho(x)}dy^2, \\
\Phi &=& \Phi_0+\frac12{\cal K}-\frac{C_1+C_2}2\rho(x). 
\end{eqnarray}

The integration constants are determined so that the solution is weakly coupled and weakly curved, 
following the criterion obtained in subsection \ref{validity}. 
The solutions for a constant $T$ are listed in Appendix \ref{trivial}.

\vspace{1cm}

\section{Non-trivial solutions} \label{non-trivial}

\vspace{5mm}

\subsection{Interpolating solutions}

\vspace{5mm}

\begin{figure}[htbp]
\begin{minipage}{.45\linewidth}
\includegraphics{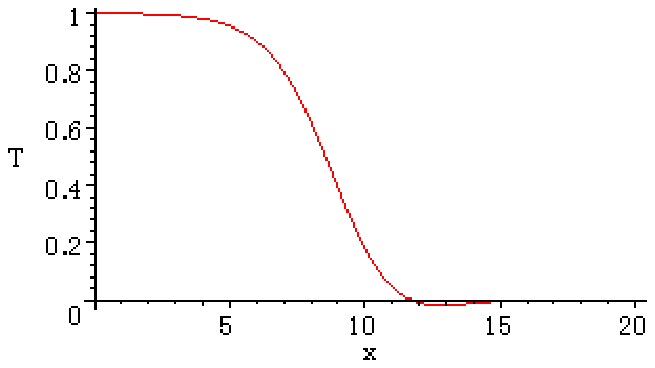}
\caption{Plot of $T(x)$}
   \label{plotT}
\end{minipage}
\begin{minipage}{.45\linewidth}
\includegraphics{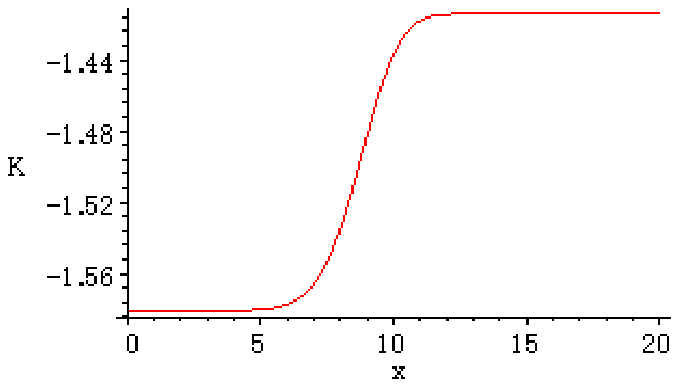}
\caption{Plot of $K(x)$}
   \label{plotK}
\end{minipage}
\end{figure}

Suppose that $V(T)$ has a (local) minimum at $T=T_{min}$. 
There is a solution in which the tachyon behaves asymptotically as 
\begin{equation}
T \to \left\{
\begin{array}{cc}
0, & (x\to+\infty) \\ T_{min}. & (x\to-\infty)
\end{array}
\right.
   \label{asymptT}
\end{equation}
The field $K$ should behave as  
\begin{equation}
K \to \left\{
\begin{array}{cc}
-\sqrt{-2V(0)}, & (x\to+\infty) \\ -\sqrt{-2V(T_{min})}. & (x\to-\infty)
\end{array}
\right.
   \label{asymptK}
\end{equation}
A plot of such a solution is shown in figure \ref{plotT} and \ref{plotK}, in which we used the 
potential $V(T)=-1-\frac12T^2+\frac14T^4$. 

Note that it is possible to consider a solution with the behavior 
\begin{equation}
\lim_{x\to-\infty}K(x)=+\sqrt{-2V(T_{min})}. 
   \label{K>0}
\end{equation}
As is shown in Appendix \ref{trivial}, the background metric 
behaves non-trivially for the $K$ with this behavior, even in the case in which the tachyon is 
constant. 
As was mentioned before, our purpose to study non-trivial solutions is to gain insights into tachyon 
condensation, especially, insights into the back-reaction of a non-zero tachyon vev to the metric. 
However, if we consider a solution obtained from a $K$ with the behavior (\ref{K>0}), then it would be 
difficult to read off the back-reaction of the tachyon vev from the $x$-dependence of the metric. 
Therefore, we focus only on solutions obtained from $T$ and $K$ which behave as (\ref{asymptT}) and 
(\ref{asymptK}). 

\vspace{3mm}

Let us determine the fields $A$, $B$ and $\Phi$ corresponding to these $T$ and $K$. 
Because $K$ approaches a negative constant in each asymptotic region, the choices of the integration 
constants for the reliable solutions are the same as in the case of the constant $K$, that is, 
$C_1=C_2=0$ in the case $H_0=0$, and $E=0$, $\rho_0\ge0$ and $C=0$ in the case $H_0\ne0$. 
The effect of the non-trivial tachyon appears as the fact that the asymptotic value of $K$ in the 
$x\to-\infty$ limit is 
different from the value in the $x\to+\infty$ limit, as is shown in (\ref{asymptK}), and also in the 
figure \ref{plotK}. 
Then, the function $\rho(x)$ behaves as 
\begin{equation}
\rho(x)\to\left\{
\begin{array}{cc}
e^{-\sqrt{-2V(0)}x}, & (x\to+\infty) \\ 
e^{-\sqrt{-2V(T_{min})}x}. & (x\to-\infty) 
\end{array}
\right.
\end{equation}
In the solutions, the dilaton gradient and the radius of the $AdS_3$ in the $x\to-\infty$ limit are 
different from those in the $x\to+\infty$ limit. 

As is shown in Appendix \ref{trivial}, there are only three kinds of solutions for a constant $K$. 
This implies that the non-trivial solutions, with the behavior (\ref{asymptT}) and 
(\ref{asymptK}), are of the three kinds:  
\begin{center}
(LD,LD), \hspace{5mm} ($AdS_3$,LD) \hspace{3mm} and \hspace{3mm} ($AdS_3$,$AdS_3$), 
\end{center}
where (A,B) indicates an interpolating solution whose background 
is asymptotically A in the $x\to-\infty$ limit, and B in the  
$x\to+\infty$ limit. 
Here LD indicates a linear dilaton. 
In terms of tachyon condensation, the existence of a solution (A,B) would suggest that B could decay 
into $A$ through tachyon condensation. 
It seems remarkable that a solution (LD,$AdS_3$) is absent. 
This would suggest that a linear dilaton cannot be obtained from an $AdS_3$ by turning on a 
tachyon vev. 
On the other hand, a linear dilaton can be deformed to an $AdS_3$. 
Therefore, the absence of (LD,$AdS_3$) seems to suggest 
that the $AdS_3$ is more stable than the linear dilaton. 

These solutions were studied in our previous paper \cite{Suyama}. 
The advantage of the analysis of this paper is that the existence of the interpolating solutions 
are shown explicitly. 

\vspace{5mm}

\subsection{Black hole solutions}

\vspace{5mm}

\begin{figure}[htbp]
\begin{minipage}{.45\linewidth}
\includegraphics{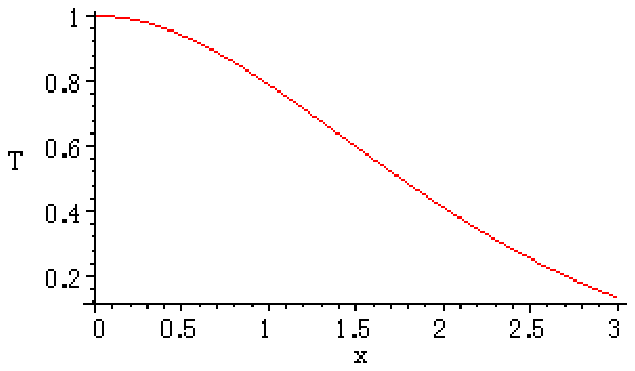}
\caption{Plot of $T(x)$}
   \label{plotT_BH}
\end{minipage}
\begin{minipage}{.45\linewidth}
\includegraphics{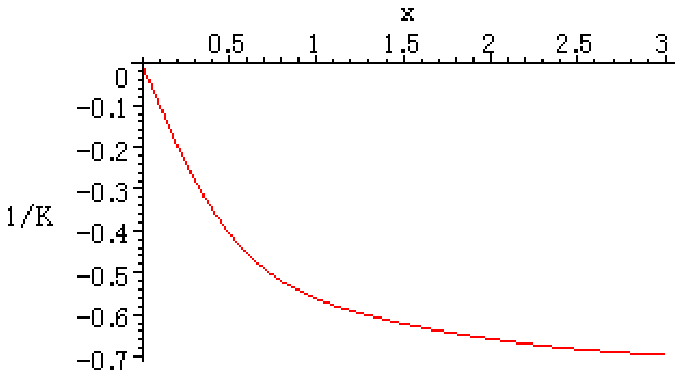}
\caption{Plot of $K(x)$}
   \label{plotK_BH}
\end{minipage}
\end{figure}

The equation (\ref{eomK}) has a solution which behaves as 
\begin{equation}
\lim_{x\to x_0}K(x) = -\infty. 
\end{equation}
Such solutions for a constant $T$ (solution III) are shown in Appendix \ref{trivial}. 
For a large negative $K$, $T$ feels a large friction, and therefore the motion of $T$ is very 
slow. 
For the limiting case $K\to-\infty$, $T$ stops at an arbitrary value, say $T_*$, irrespective of 
whether the value $T=T_*$ corresponds to an extremum of $V(T)$ or not. 
An example of such a solution is plotted in figures \ref{plotT_BH} and \ref{plotK_BH}. 

For this plot, we employed the potential $V(T)=-1-\frac12T^2$, and we chose $x_0=0$. 

To determine the condition for the reliable solutions, let us investigate the behavior of a solution 
for $\Phi$ near $x=0$. 
Assume the behavior of $T(x)$ as 
\begin{equation}
T(x) = \sum_{n=0}^\infty t_nx^n, 
\end{equation}
where $t_1=0$. 
Then $K$ behaves as 
\begin{equation}
K(x) = -\frac1x+\frac23V(t_0)x+O(x^3). 
   \label{behaveK}
\end{equation}
Therefore, we obtain $\rho(x)\sim\log x$ in the $x\to0$ limit. 

In the case $H_0=0$, $\Phi'$ is finite if $C_1+C_2=-1$, and in this case 
\begin{equation}
\Phi'(x) = \frac12V(t_0)x+O(x^2). 
\end{equation}
Because $\Phi'$ vanishes at $x=0$, either $A'$ or $B'$ may diverge. 
Suppose that $A'$ diverges, which corresponds to the choice $C_1=-1$ and $C_2=0$. 
Then the metric behaves near $x=0$ as 
\begin{equation}
ds^2 \sim -x^2dt^2+dx^2+dy^2. 
   \label{Rindler}
\end{equation}
The RHS is the Rindler metric with the flat $y$-direction. 
The behavior of this metric indicates that there is an event horizon at $x=0$. 
Because $B$ is constant everywhere in this solution, this black hole is a two-dimensional one. 

In the case $H_0\ne0$, the condition for $\Phi'$ to be finite implies $\sqrt{2E}=1$. 
In this case, $\Phi'$ behaves as 
\begin{equation}
\Phi'(x) = \left(\frac12V(t_0)+2e^{-2\rho_0}\right)x+O(x^2). 
\end{equation}
As in the case $H_0=0$, either $A'$ or $B'$ may diverge at $x=0$. 
For the choice $C=-1$, $A'$ and $B'$ behave as 
\begin{eqnarray}
A'(x) &=& -\frac1x-\left(\frac13V(t_0)+e^{-2\rho_0}\right)x+O(x^3), 
   \label{behaveA} \\
B'(x) &=& -e^{-2\rho_0}x+O(x^3). 
   \label{behaveB}
\end{eqnarray}
Therefore, the behavior of the metric near $x=0$ is again that of the Rindler metric (\ref{Rindler}). 
In this case, because $B$ is non-trivial, this black hole is a three-dimensional one. 
The constant $\rho_0$ determines the behavior in the $x\to+\infty$ limit, as in the case of the 
interpolating solutions discussed in the previous subsection. 
The solution is asymptotically $AdS_3$ for the choice $\rho_0=0$, while for the choice $\rho_0>0$ 
the solution approaches the linear dilaton. 

\vspace{3mm}

The existence of these solutions indicates that a local tachyon condensation might form a black hole, 
and the region where the vev of the tachyon grows might be hidden behind the horizon. 
It is remarkable that these black hole solutions can be realized even in the case $V(0)=0$ for which 
any solution with a constant $T$ does not have a horizon, and therefore, the appearance of the horizon 
is due to the non-trivial behavior of the tachyon vev. 
It is also remarkable that $t_0$ can be arbitrarily small, and therefore the fields are slowly varying 
irrespective to the curvature of the potential $V(T)$ at $T=0$. 
These remarks suggest that these solutions could be reliable ones also for a critical string theory 
with a tachyon whose mass squared is of the string scale, for example, Type 0 string. 
A further interesting point of these solutions is that they exist for the potential $V(T)$ 
which is {\it unbounded from below}. 

Note that the behavior of the metric (\ref{Rindler}) 
does not directly imply the presence of a horizon, 
because the existence of the horizon depends on the global structure of the spacetime. 
However, because our time coordinate $t$ is the usual one in the asymptotic region $x\to+\infty$, 
and therefore the surface $x=0$ corresponds to $t=\pm\infty$ in the viewpoint of the asymptotic 
observer, it is natural to expect that $x=0$ is a horizon. 

\vspace{5mm}

\subsection{Capped geometry}

\vspace{5mm}

Due to the invariance of the equations of motion under the exchange $A\leftrightarrow B$, there are 
solutions in which $B'$, instead of $A'$ diverges. 
The metric of the solution behaves near $x=0$ as 
\begin{equation}
ds^2 \sim -dt^2+dx^2+x^2dy^2, 
\end{equation}
in which the $y$-direction, if compactified, is pinched off at $x=0$. 
As in the case of the black hole solution, the non-trivial part of the spacetime is two-dimensional 
for the case $H_0=0$, and three-dimensional for the case $H_0\ne0$. 

In the region $x<0$, the tachyon vev is supposed to become large if the back-reaction is ignored, 
and therefore, these solutions 
would indicate the existence of a mechanism by which a region with a large tachyon vev is 
eliminated, similar to that studied in \cite{Silverstein}. 
It is interesting to notice that in our case a cylindrical part of the spacetime is pinched off by a 
condensation of a tachyon with {\it winding number zero}. 

\vspace{5mm}

\subsection{Constraint}

\vspace{5mm}

It should be checked whether the above solutions satisfy the constraint (\ref{constraint}). 
Let us define 
\begin{equation}
F = K^2-(A')^2-(B')^2-(T')^2+2V(T)+\frac12H_0^2e^{4\Phi}. 
\end{equation}
It is easy to check that $F$ vanishes for the linear dilaton and the $AdS_3$. 
Using the equations of motion, it can be shown that 
\begin{equation}
\frac{dF}{dx} = 2KF, 
\end{equation}
and therefore, we obtain 
\begin{equation}
F = C_0e^{2{\cal K}}. 
\end{equation}

First, consider the interpolating solutions. 
The solutions approach the linear dilaton or the $AdS_3$ asymptotically, and thus $F$ vanishes in 
the $x\to\pm\infty$ limit. 
This is possible only for the choice $C_0=0$, implying that the solutions satisfy the constraint. 

Next, consider the black hole solutions and the capped geometry solutions. 
For these solutions $e^{2{\cal K}}$ diverges at $x=0$. 
However, it can be shown from the behavior (\ref{behaveK}), (\ref{behaveA}) and (\ref{behaveB}) that 
the divergences in $F$ are canceled. 
This implies $C_0=0$, that is, the constraint is satisfied.

\vspace{1cm}

\section{Masses of the solutions} \label{mass}

\vspace{5mm}

In this section, we determine the masses of the solutions obtained in the previous section to discuss 
their stability. 

We have considered the metric
\begin{equation}
ds^2_E = e^{\alpha\Phi}\left[-e^{-2A(x)}dt^2+dx^2+e^{-2B(x)}dy^2+dz^idz^i\right]
\end{equation}
in the Einstein frame, 
where $i$ runs from 1 to $D-3$, and in our case we have $\alpha=-\frac4{D-2}$. 

Following \cite{HH}, we define the mass $M(A,B,\Phi)$ of a solution as  
\begin{equation}
M(A,B,\Phi) = -\frac1{\kappa^2}\int_{S^\infty}NK_{e}
 -\frac1{24\kappa^2}\int_{S^\infty}e^{\frac8{D-2}\Phi}H^{xyt}B_{yt}.
   \label{ADM}
\end{equation}
Here we chose the spatial hypersurface $\Sigma$ as a hypersurface with a 
constant time, and $S^\infty$ is 
a boundary of $\Sigma$ specified by $x=+\infty$. 
The quantities $N$ and $K_{e}$ are given as follows, 
\begin{eqnarray}
N &=& e^{\frac{\alpha}2\Phi-A}, \\
K_{e} &=& \frac12e^{-\frac\alpha2}[\alpha(D-2)\Phi'-2B']. 
\end{eqnarray}
By substituting them into (\ref{ADM}), the mass is given as 
\begin{equation}
M(A,B,\Phi) = -\frac{\cal V}{2\kappa^2}\Bigl[ 
 e^{-A-B+\frac{D-2}2\alpha\Phi}[(D-2)\alpha\Phi'-2B'] 
 +\frac1{12}H_0e^{A+(\frac{D-2}2\alpha+2)\Phi}B_{yt}\Bigr]\Big|_{x\to+\infty}, 
   \label{massformula}
\end{equation}
where ${\cal V}=\int dyd^{D-3}z$ is the volume of the boundary $S^\infty$. 
If there is a horizon in the solution, then the contribution from the horizon should also be taken into 
account. 

The quantity (\ref{massformula}) may diverge, and thus the mass of a solution is usually defined by 
the difference from the mass $M(A_0,B_0,\Phi_0)$ of a reference solution. 
One possible 
choice of the reference solution is either a linear dilaton or an $AdS_3$, according to the 
asymptotic behavior of the solution in question. 
However, this choice does not work for non-trivial solutions (solutions with non-trivial tachyon 
profiles). 
Recall that the tachyon $T$ may oscillate in the $x\to+\infty$ limit. 
In the case of an asymptotically $AdS_3$ solution, the oscillation occurs exactly when the mass 
squared of the tachyon is below the BF bound, and therefore, the oscillating behavior is relevant 
for cases we are interested in. 
Due to this oscillation of $T$, and thus the other fields, the quantity 
$M(A,B,\Phi)-M(A_0,B_0,\Phi_0)$ is not well-defined. 

Although the mass relative to trivial solutions are not well-defined, it is possible to define a 
relative mass between two non-trivial solutions. 
As mentioned before, there is always a pair of solutions which are related to each other by exchanging 
$A$ and $B$. 
The mass difference $\Delta M=M(A,B,\Phi)-M(B,A,\Phi)$ between these solutions 
is then well-defined, because the oscillatory behavior cancels. 
The mass difference $\Delta M$ so defined is non-zero only for the cases of the 
black hole solutions and the capped geometry solutions. 
Hence we concentrate on these cases. 

Let $\Delta M_g$ denote the first term of the RHS of (\ref{massformula}). 
This is evaluated to be 
\begin{eqnarray}
\Delta M_g &=& -\frac{\cal V}{\kappa^2}e^{-A-B-2\Phi}(A'-B')\Big|_{x\to+\infty} \nonumber \\
&=& -\frac{\cal V}{\kappa^2}C, 
\end{eqnarray}
where $C$ is the integration constant in (\ref{intA-B}). 
The remaining term $\Delta M_B$ is evaluated as follows. 
Because the component $H^{txy}$ is invariant under the exchange of $A$ and $B$, the component 
$B_{yt}$ is also invariant. 
Therefore, 
\begin{eqnarray}
\Delta M_B &\propto& (e^A-e^B)B_{yt}\Big|_{x\to+\infty} \nonumber \\
&=& e^{\frac12(A+B)}(A-B)B_{yt}\Big|_{x\to+\infty} \nonumber \\
&\sim& \lim_{x\to+\infty}\left\{
\begin{array}{cc}
e^{2K(+\infty)x}, & (\mbox{LD}) \\
e^{\frac12K(+\infty)x}. & (AdS_3) 
\end{array}
\right. 
\end{eqnarray}
Because $\lim_{x\to+\infty}K(x)$ is negative in our solutions, the contribution $\Delta M_B$ vanishes. 

For the mass of the black hole solution, the contribution from the horizon must be taken into 
account. 
Because the lapse $N$ vanishes at the horizon, 
the contribution to $\Delta M_g$ from the horizon vanishes. 
The contribution $\Delta M_{B,h}$ to $\Delta M_B$ from the horizon 
behaves as $\Delta M_{B,h}\sim \sqrt{x}$, where the horizon is at $x=0$, 
and therefore this vanishes. 

Now we obtained the following formula for the mass difference of the solutions, 
\begin{equation}
\Delta M = -\frac{\cal V}{\kappa^2}C. 
\end{equation}
Recall that the constant 
$C$ is negative when $\Delta M$ is the black hole mass minus the mass of the capped 
geometry. 
Therefore, the black hole is heavier than the capped geometry. 
This implies that, if there is some physical process which deforms the black hole geometry into the 
capped geometry, 
then the latter is likely to be realized after tachyon condensation, unless the capped geometry 
decays further. 
It is very interesting to clarify whether there exists such a process.

\vspace{1cm}

\section{Discussion} \label{discuss}

\vspace{5mm}

We have investigated classical solutions of the effective action (\ref{EA}) with $f(T)=0$. 
Employing the ansatz presented in section \ref{EOM}, 
we could find all solutions systematically, provided that 
solutions of the equations (\ref{eomT}) and (\ref{eomK}) were given. 
If there is a minimum in the potential $V(T)$, then there are solutions which interpolate a 
linear dilaton or an 
$AdS_3$ corresponding to the minimum of $V(T)$ and those corresponding to the maximum. 
In addition, there are black hole solutions and capped geometry solutions. 
It is quite remarkable that the latter solutions are acceptable as solutions of the 
low energy effective 
theory, even in the case in which the curvature of the potential at $T=0$ is of the order 
of the string scale. 
It is therefore expected that the solutions could have some implications to a condensation of  
a tachyon whose mass squared is of the order of the string scale. 

It is very interesting to determine what kind of a background is realized inside the horizon of the 
black hole solution.\footnote{I would like to thank S.Rey for valuable discussions} 
Because we only know the metric outside the horizon, it is difficult to determine 
how to extend our metric beyond the horizon. 
As for the usual BTZ black hole \cite{BTZ}, 
it would be possible to attach a solution in which the role of $t$ 
and $x$ are exchanged, and this extension would produce 
a singular geometry inside the horizon. 
Interestingly, there is another way of extension which enables us to obtain a solution which is 
non-singular everywhere. 
It can be shown that our black hole solutions are invariant under the reflection $x\to-x$. 
Therefore, we can simply extend our solution to the region $x<0$ by defining, 
for example, $T(x):=T(-x)$. 
The resulting solution has infinitely many asymptotic regions, similarly to the D3-brane 
solution \cite{D3}, and there is no place where the vev of the tachyon grows without bound. 
This solution might suggest the existence of the following process of tachyon condensation. 
As the vev of the tachyon grows at some region, this region is replaced with a geometry with another 
asymptotic region, and the tachyon vev is finite in the entire spacetime. 
There are also the other solutions, the capped geometry solutions. 
In such a solution, the region with a large tachyon vev is simply eliminated. 
These solutions seem to suggest that there might be a mechanism to avoid growing the tachyon vev 
without bound. 

It should be noted that our solutions are asymptotically {\it tachyonic}, that is, $T$ approaches zero 
in the $x\to+\infty$ limit. 
This means that our solutions must be unstable, and they decay further. 
(It might be possible that the region $x\to-\infty$ is rather stable compared with the region 
$x\to+\infty$.) 
A possible process for a further decay would be to form more black holes. 
The final state of such a process would be a very complicated one, and it is still unclear what kind 
of string theory is realized after tachyon condensation. 

It may be interesting to study classical solutions of an action which has a $p$-form field 
strength, as a simple generalization of our analysis. 
Let us consider the action 
\begin{equation}
S = \frac1{2\kappa^2}\int d^Dx\sqrt{-g}\ e^{-2\Phi}\left[ R+4(\nabla\Phi)^2-\frac12e^{\alpha\Phi}
 |F_p|^2-(\nabla T)^2-2V(T) \right], 
\end{equation}
where $\alpha$ is a constant. 
We followed the notation of \cite{Pol} for the kinetic term of the form field. 
Let us consider the ansatz
\begin{equation}
ds^2 = -e^{-2A(x)}dt^2+dx^2+e^{-2B(x)}dy^mdy^m+e^{-2C(x)}dx^adz^a, 
\end{equation}
for the metric, and assume that the dilaton and the tachyon are functions of $x$. 
Then the equations of motion reduce to 
\begin{eqnarray}
A''-KA'+\left(\frac\alpha8-\frac12\right)e^{\alpha\Phi}|F_p|^2 &=& 0, \\
B''-KB'+\left(\frac\alpha8-\frac12\right)e^{\alpha\Phi}|F_p|^2 &=& 0, \\
C''-KC'+\frac\alpha8e^{\alpha\Phi}|F_p|^2 &=& 0, \\
\Phi''-K\Phi'-V(T)+\left(\frac{p-1}4-\frac\alpha{16}(D-2)\right)e^{\alpha\Phi}|F_p|^2 &=& 0, \\
K'-K^2-2V(T)+\frac\alpha8e^{\alpha\Phi}|F_p|^2 &=& 0, \\
T''-KT'-V'(T) &=& 0, 
\end{eqnarray}
where 
\begin{equation}
K = A'+(p-2)B'+(D-p)C'+2\Phi'. 
\end{equation}
The fields $T$ and $K$ decouple from the other fields only when $\alpha=0$. 
In this case, the classification of classical solutions goes similarly to our case ($p=3$), but the 
results for the case with $p\ne3$ and $\alpha=0$ have nothing to do with string theory. 

It is very interesting to consider a more general solutions, for example, a solution describing a 
time evolution of an inhomogeneous tachyon profile. 
It is also interesting to know whether the non-trivial tachyon solutions are more stable than the 
trivial tachyonic vacuum. 
We hope that 
more detailed investigations on classical solutions would provide us with much insight into 
closed string 
tachyon condensation.

\vspace{2cm}

\begin{flushleft}
{\Large \bf Acknowledgments}
\end{flushleft}

\vspace{5mm}

I would like to thank K.Furuuchi, J.Nishimura, S.Rey and T.Takayanagi for valuable discussions. 
This work was supported in part by JSPS Research Fellowships for Young Scientists.

\appendix

\vspace{2cm}

{\bf \LARGE Appendices}

\vspace{1cm}

\section{Effective action} \label{action}

\vspace{5mm}

In this Appendix, we show that the effective action of the string theory mentioned in section 
\ref{EOM} has the form (\ref{EA}). 

The string theory we consider is assumed to be compactified on $M$ where $M$ is a 
compact manifold, and the tachyon corresponds to a relevant operator of a CFT describing $M$. 
We assume that all massless fields as well as the tachyon varies slowly in the target spacetime, which 
enables us to ignore all terms with more than two derivatives. 
The most general action for the metric, the B-field, the dilaton and the tachyon is 
\begin{eqnarray}
S 
&=& \frac1{2\kappa^2}\int d^Dx\sqrt{-g}\ e^{-2\Phi}f_1(T)\Bigl[ R+4f_2(T)(\nabla\Phi)^2 
 -\frac1{12}f_3(T)H^2  \nonumber \\
& & +f_4(T)\nabla\Phi\cdot\nabla T-f_5(T)(\nabla T)^2-2V(T) \Bigr]. 
   \label{stringframe}
\end{eqnarray}
Here it is also assumed that a constant 
shift of $\Phi$ is always absorbed by $\kappa$, as should be the case. 

Some of $f_i(T)$ can be chosen to simple functions by field redefinitions. 
To show this, it may be convenient to consider the effective action in the Einstein frame, 
\begin{eqnarray}
S 
&=& \frac1{2\kappa^2}\int d^Dx\sqrt{-g_E}\ f_1(T)\Bigl[ R_E-\frac4{D-2}\tilde{f}_2(T)(\nabla_E\Phi)^2 
 -\frac1{12}f_3(T)e^{-\frac{8\Phi}{D-2}}H^2  \nonumber \\
& & +\tilde{f}_4(T)\nabla_E\Phi\cdot\nabla_E T-f_5(T)(\nabla_ET)^2-2e^{\frac{4\Phi}{D-2}}V(T) \Bigr], 
   \label{generic action}
\end{eqnarray}
where 
\begin{eqnarray}
\tilde{f}_2(T) &=& D-1-(D-2)f_2(T), \\
\tilde{f}_4(T) &=& f_4(T)+4\frac{D-1}{D-2}\frac{f_1'(T)}{f_1(T)}. 
\end{eqnarray}
The subscript $E$ indicates that they are constructed in terms of the Einstein metric $g_E$. 
It is possible to set $f_1(T)=1$ by a $T$-dependent Weyl rescaling $g_E\to e^{2\omega(T)}g_E$,
and $\tilde{f}_4(T)=0$ by a $T$-dependent shift $\Phi\to\Phi+\alpha(T)$, where $\alpha(T)$ and 
$\omega(T)$ are determined by 
\begin{eqnarray}
\partial_T\alpha(T) &=& \frac{D-2}8\frac{\tilde{f}_4(T)}{\tilde{f}_2(T)}, \\
(D-2)\omega(T) &=& -\log f_1(T). 
\end{eqnarray}
The kinetic term of $T$ can be normalized by a suitable redefinition of $T$, that is, we can have 
$f_5(T)=1$ without loss of generality. 
Note that the string frame action (\ref{stringframe}) providing 
$f_1(T)=1,\ \tilde{f}_4(T)=0,\ f_5(T)=1$ is the action 
with $f_1(T)=1,\ f_4(T)=0$ and $f_5(T)=1$. 
In summary, there are two undetermined functions of $T$, $f_2(T)$ and $f_3(T)$, 
in addition to $V(T)$, which cannot be 
fixed simply by field redefinitions. 

To restrict the form of the action further, we need an assumption on the dynamics of this system. 
We assume the existence of the correspondence between 
the equations of motion of the action (\ref{stringframe}) and the beta-functionals for the conformal 
invariance of the underlying worldsheet theory. 
The action of the worldsheet theory is 
\begin{eqnarray}
S_{ws} &=& \frac1{4\pi\alpha'}\int d^2\sigma\sqrt{h}\Bigl[ \partial_a X^\mu\partial_b X^\nu\bigl( 
 h^{ab}g_{\mu\nu}(X)+i\epsilon^{ab}B_{\mu\nu}(X)\bigr)+\alpha'R^{(2)}\Phi(X) \Bigr] \nonumber \\
& & +S_M(Y)+\int d^2\sigma\sqrt{h}\ T(X)V(Y). 
   \label{worldsheet}
\end{eqnarray}
We followed the notation of \cite{Pol}. 
The action $S_M(Y)$ is that of a CFT 
describing $M$, for which the dynamical variables are collectively denoted by $Y$. 
The operator $V(Y)$ is a vertex operator of the CFT corresponding to a tachyon. 
The effective action (\ref{stringframe}) should be derived from (\ref{worldsheet}). 
Suppose that $T(X)$ is a constant. 
Then the non-compact part and the $M$ part of the worldsheet theory (\ref{worldsheet}) 
decouple from each other. 
In this case, the beta-functional $\beta_{\mu\nu}$ 
for $g_{\mu\nu}$, for example, does not depend on $T$ at all. 
This implies that the beta-functional $\beta_{\mu\nu}$ for a generic $T(X)$ should be 
\begin{equation}
\beta_{\mu\nu} = R_{\mu\nu}+2\nabla_\mu\nabla_\nu\Phi-\frac14H_{\mu\rho\lambda}H_\nu{}^{\rho\lambda}
 +\mbox{ derivatives of }T, 
\end{equation}
at the leading order in $\alpha'$. 
We require that $\beta_{\mu\nu}$ is equal to a linear combination of 
$\frac{\delta S}{\delta g^{\mu\nu}}$ and $\frac{\delta S}{\delta \Phi}$, which is possible 
only if $f_1(T)=f_2(T)=f_3(T)=1$. 
We can choose $f_5(T)=1$ as before. 
Therefore, the only undetermined functions of $T$ are $f_4(T)$ and $V(T)$. 
The action (\ref{EA}) can be obtained by a field redefinition.

\vspace{1cm}

\section{Classical solutions with constant $T$} \label{trivial}

\vspace{5mm}

Assuming that $T$ is constant, the equation (\ref{eomK}) can be solved exactly. 
All solutions to this equation are 
\begin{equation}
K(x) = \left\{ 
\begin{array}{cc}
\pm v, & \mbox{(I)} \\ -v\tanh(v(x-x_0)), & \mbox{(II)} \\ -v\coth(v(x-x_0)), & \mbox{(III)}
\end{array}
\right.
\end{equation}
where $v^2=-2V(T)\ge 0$, and $x_0$ is an integration constant. 
The solution III is singular at $x=x_0$, and therefore, the range of $x$ is either 
$-\infty<x\le x_0$ or $x_0\le x<+\infty$. 
Because the equations of motion (\ref{eomT})-(\ref{A-B}) are invariant under the reflection $x\to-x$ 
and $K\to-K$, 
these two choices for the range of $x$ are related to each other. 
Similarly, the solution $K=v$ is related to the solution $K=-v$. 

Note that no reliable solutions are obtained if a potential with the property $V(T)>0$ is considered. 

\vspace{5mm}

\subsection{Solution I}

\vspace{5mm}

We consider the solution $K=-v\le0$. 

First, consider the $H_0=0$ case. 
The general solution for $A$, $B$ and $\Phi$ is  
\begin{eqnarray}
A'(x) &=& C_1e^{-vx}, \\
B'(x) &=& C_2e^{-vx}, \\
\Phi'(x) &=& -\frac v2-\frac{C_1+C_2}2e^{-vx}. 
\end{eqnarray}
The finiteness of $\Phi'$ is realized by choosing $C_1+C_2=0$. 
Because it is not allowed for $A'$ and $B'$ to diverge simultaneously, we have to choose $C_1=-C_2=0$. 
Therefore, the only reliable solution is the linear dilaton solution in the flat spacetime. 

\vspace{3mm}

In the case $H_0\ne0$, the general solution is obtained by substituting  
\begin{equation}
\rho(x) = -\frac1ve^{-vx}, 
\end{equation}
into the equations (\ref{solnA'})-(\ref{solnPhi'}). 
This solution is reliable only for the choice $E=0$, $\rho_0\ge0$ and $C=0$. 
In this case, we obtain 
\begin{eqnarray}
A'(x) &=& -\frac v2\frac{e^{-vx}}{\rho_0v+e^{-vx}}\ =\ B'(x), \\
\Phi'(x) &=& -\frac v2\left(1-\frac{e^{-vx}}{\rho_0v+e^{-vx}}\right). 
\end{eqnarray}
The metric corresponding to this solution is 
\begin{equation}
ds^2 = \frac1{\rho_0v+e^{-vx}}(-dt^2+dy^2)+dx^2. 
\end{equation}
The solution with $\rho_0=0$ is the $AdS_3$ with the constant dilaton. 
In the case $\rho_0>0$, the solution interpolates the linear dilaton and the center of the $AdS_3$. 
Note that there is no solution interpolating the linear dilaton and the {\it boundary} of the $AdS_3$. 

\vspace{5mm}

\subsection{Solution II}

\vspace{5mm}

For the solution II, we have 
\begin{equation}
e^{{\cal K}(x)} = \frac1{\cosh(v(x-x_0))}.
\end{equation}

\vspace{3mm}

In the case $H_0=0$, it is obvious that the solution with an arbitrary choice of the integration 
constants is reliable. 
In the $x\to\pm\infty$ limits, the solution approaches the linear dilaton with the dilaton gradient 
\begin{equation}
\lim_{x\to\pm\infty}\Phi'(x) = \mp\frac v2. 
\end{equation}
This shows that both the asymptotic regions are weakly coupled. 

\vspace{3mm}

In the case $H_0\ne0$, the function $\rho(x)$ is given as 
\begin{equation}
\rho(x) = \frac1v\mbox{arctan}\left[ \sinh(v(x-x_0))\right]. 
\end{equation}
The range of $\rho$ is $-\frac \pi{2v}<\rho<\frac\pi{2v}$. 
If we choose $|\rho_0|<\frac\pi{2v}$, 
then $\Phi'$ diverges at a finite $x$, resulting in a singular solution. 
Because the case $\rho_0\le-\frac\pi{2v}$ is related to the case $\rho_0\ge\frac\pi{2v}$ by the 
reflection of $x$, we investigate only the latter case. 

In the limit $x\to-\infty$, it can be shown that the solution approaches the linear dilaton 
generically, while for the negative $E$ satisfying
\begin{equation}
\sqrt{2|E|}\left(\frac\pi{2v}+\rho_0\right) = \pi, 
\end{equation}
the solution approaches the $AdS_3$. 
The behavior of the solution in the limit $x\to+\infty$ depends on the choice of $\rho_0$. 
For the choice $\rho_0>\frac\pi{2v}$, on the one hand, 
the solution also approaches the other linear dilaton. 
The sign of the dilaton gradient in the limit $x\to+\infty$ is opposite to that in the limit 
$x\to-\infty$, and therefore, the string coupling is small everywhere. 
For the choice $\rho_0=\frac\pi{2v}$, on the other hand, the solution approaches the boundary of 
the $AdS_3$. 

In summary, there are interpolating solutions which approach in the $x\to\pm\infty$ limits either 
the linear dilaton or the $AdS_3$. 
Any combination of the asymptotic behaviors is possible, and thus there are 
four kinds of solutions. 
It should be noted that there always appear the weak coupling region of the linear dilaton, and 
the boundary of the $AdS_3$. 

\vspace{5mm}

\subsection{Solution III}

\vspace{5mm}

For definiteness, we choose the range $x_0\le x<+\infty$. 
For this solution, we have 
\begin{equation}
e^{{\cal K}(x)} = \frac1{\sinh(v(x-x_0))}. 
\end{equation}

\vspace{3mm}

In the case $H_0=0$, the finiteness of $\Phi'$ at $x=x_0$ implies $C_1+C_2=-v$. 
Because $A'B'$ must be also finite, either $C_1$ or $C_2$ must be zero. 
Therefore, there are two reliable solutions. 
One solution, for the choice $C_1=-v$, is 
\begin{eqnarray}
A'(x) &=& -\frac v{\sinh(v(x-x_0))}, \\
B'(x) &=& 0, \\
\Phi'(x) &=& -v\tanh\frac v2(x-x_0), 
\end{eqnarray}
which corresponding to the metric 
\begin{equation}
ds^2 = -\tanh^2\frac v2(x-x_0)dt^2+dx^2+dy^2. 
\end{equation}
This is Witten's black hole \cite{WittenBH}. 
The other solution, for the choice $C_2=-v$, is 
\begin{eqnarray}
A'(x) &=& 0, \\
B'(x) &=& -\frac v{\sinh(v(x-x_0))}, \\
\Phi'(x) &=& -v\tanh\frac v2(x-x_0), 
\end{eqnarray}
which corresponds to the Euclidean cigar geometry 
\begin{equation}
ds^2 = -dt^2+dx^2+\tanh^2\frac v2(x-x_0)dy^2. 
\end{equation}
There is no singularity at $x=x_0$ if the $y$-coordinate is compactified with the radius $\frac2v$. 

\vspace{3mm}

Next, we consider the case $H_0\ne0$. 
In this case, $\rho(x)$ is given as 
\begin{equation}
\rho(x) = \frac1v\log\left[\tanh\frac v2(x-x_0)\right]. 
\end{equation}
The finiteness of $\Phi'$ at $x=x_0$ implies $\sqrt{2E}=v$, and the finiteness of $A'B'$ implies 
$C=\pm v$. 
Therefore, also in this case, there are two reliable solutions. 
One solution, for the choice $C=-v$ is 
\begin{eqnarray}
A'(x) &=& \frac v{\sinh(v(x-x_0))}\frac{1}{a^2\tanh^2\frac v2(x-x_0)-1}, \\
B'(x) &=& \frac v{\sinh(v(x-x_0))}\frac{a^2\tanh^2\frac v2(x-x_0)}{a^2\tanh^2\frac v2(x-x_0)-1}, \\
\Phi'(x) &=& -\frac v{2\sinh(v(x-x_0))}\left[ \cosh(v(x-x_0))
 +\frac{a^2\tanh^2\frac v2(x-x_0)+1}{a^2\tanh^2\frac v2(x-x_0)-1} \right], 
\end{eqnarray}
where $a=e^{-\rho_0v}$. 
The other solution, for the choice $C=+v$, is obtained from this solution by exchanging $A$ and $B$. 
It can be understood from the above expression that the finiteness of the solution also implies 
$a^2\le1$, that is, $\rho_0\ge0$. 

The solution has a simple form for $a=1$. 
In this case, $\Phi'$ exactly vanishes, and the metric, for $C=-v$, is 
\begin{equation}
ds^2 = -\sinh^2\frac v2(x-x_0)dt^2+dx^2+\cosh^2\frac v2(x-x_0)dy^2. 
\end{equation}
Defining the new coordinate as 
\begin{equation}
r=\cosh\frac v2(x-x_0), 
\end{equation}
the above metric is rewritten as 
\begin{equation}
ds^2 = -\left(\frac{r^2}{l^2}-M\right)dt^2+\frac{dr^2}{\frac{r^2}{l^2}-M}+r^2dy^2, 
\end{equation}
where $l^2=\frac 4{v^2}$, and $M=\frac {v^2}4$. 
This metric has the same form with that of the non-rotating BTZ black hole \cite{BTZ}. 
By compactifying the $y$-coordinate with the radius $R$, 
the solution indeed describes the BTZ black hole. 
The mass of the black hole is then $MR^2$. 

The metric of the solution with $C=+v$ is 
\begin{equation}
ds^2 = -\cosh^2\frac v2(x-x_0)dt^2+dx^2+\sinh^2\frac v2(x-x_0)dy^2. 
\end{equation}
Suppose that the $y$-coordinate is compactified with the radius $R$. 
Then the metric describes a spacetime whose spatial section is a cylinder pinched off at $x=x_0$. 
There is no conical singularity if $R=\frac 2v$. 

The solution with the choice $0<a<1$ has a rather complicated expression. 
However, the behavior around $x=x_0$ is almost the same as that of the solution with $a=1$, 
because $1-a^2$ appears only in higher order terms of $x-x_0$. 
A qualitative difference from the solution with $a=1$ is that the solution with $0<a<1$ approaches 
asymptotically the linear dilaton in the $x\to+\infty$ limit, while the BTZ black hole is 
asymptotically $AdS_3$.

\end{document}